\documentstyle[aps,preprint]{revtex}
\tighten

\newcommand{\ls}{$l^*$}

\begin{document}
\preprint{UMDGR 98-19}
\title{GRAVITATIONAL GEONS REVISITED}
\author{Paul R. Anderson}
\address{Department of Physics\\Wake Forest University\\P.O. Box
7507\\Winston-Salem, NC  27109}
\author{Dieter R. Brill}
\address{Department of Physics\\University of Maryland\\College
Park, MD  20742}
\maketitle
\begin{abstract}

     A careful analysis of the gravitational geon solution found by
Brill and Hartle is made.  The gravitational wave expansion they
used is shown to be consistent and to result in a gauge invariant
wave equation.  It also results in a gauge invariant effective
stress-energy tensor for the gravitational waves provided that a
generalized definition of a gauge transformation is used.  To
leading order this gauge transformation is the same as the usual
one for gravitational waves.  It is shown that the geon solution is
a self-consistent solution to Einstein's equations and that, to
leading order, the equations describing the geometry of the
gravitational geon are identical to those derived by Wheeler for
the electromagnetic geon. An appendix provides an existence proof for
geon solutions to these equations.

\end{abstract}
\pacs{98.8}
\section{Introduction}

     Brill and Hartle \cite{BH}, BH, 
developed a very useful method for finding approximate solutions to Einstein's
equations that correspond to high frequency gravitational waves
propagating in a background geometry, which is created by the
average stress-energy of the waves themselves.  In their paper 
they applied this
method to the case of a static spherically symmetric background
geometry and found that gravitational waves can remain confined
in a region for a time much longer than the region's light-crossing
time.  This so-called {\em gravitational geon} is
generated by a large number of high frequency, small amplitude
gravitational waves.  The time average of the curvature due to
these waves creates the background geometry of the geon, and this
background geometry traps the waves for a long time in a
region of space called the ``active'' region.\footnote{Gravitational 
geons are analogous to the original
electromagnetic geons of Wheeler \cite{Wheeler}, which are
virtual gravitationally bound states of electromagnetic energy.
As such, geons have a finite lifetime and are not true non-radiative
solutions. Gibbons and Stewart \cite{GS} have shown that
non-radiative geons, or other exactly periodic solutions,
cannot exist in Einstein's theory.}  
The BH solution is important because it serves as an example in which the
gravitational field 
both creates and responds to its own effective stress-energy.  It
is also an example of a nontrivial (approximate) solution to the
vacuum Einstein equations that has no curvature singularities.

There are three
reasons for analyzing the solution found by BH in more
detail than was done in their original paper.  The first is that BH 
did not investigate
the question, is the stress-energy tensor they used conserved
or gauge invariant?  To have a self-consistent set of equations it
is necessary that the effective stress-energy tensor have these
properties.
Second, the validity of the geon solution found by BH has been
questioned by Cooperstock, Faraoni, and Perry \cite{C1,C2}. 
They claim that the thin shell approximation used by BH to describe
the active region is invalid, and that gravitational geons
due to high frequency, large angular momentum waves do not exist.
The third reason for analyzing the BH solution in more detail
is that, because they used a thin shell approximation, BH did
not determine the form of the geometry within the active region. 
The active region is the only region  where the
gravitational waves have a significant size, and it is the
region where spacetime is most strongly curved. It is
clearly important to know the geometry of the active region, 
if one wishes to learn anything about the details of the geon
solution. 

     In this paper a thorough analysis of the geon solution found
by BH is presented.  A self-consistent expansion of the metric and
curvature tensors is found.  Using results recently obtained
regarding effective stress-energy tensors for gravitational 
waves\cite{stressten},
it is argued that, to leading order, the wave equation and
stress-energy tensor are gauge invariant.  It is also shown that the
stress-energy tensor is conserved to leading order with respect to
the background geometry.
The wave and backreaction equations are explicitly derived in
the high frequency and large angular momentum limits.  It is shown
that in and near the active region these equations can be cast in
a form which is mathematically identical to the equations derived
by Wheeler \cite{Wheeler} for the electromagnetic geon.  Thus, to
leading order, the background geometry of the gravitational geon
found by BH is identical to that of the electromagnetic geon found
by Wheeler.\footnote{It is already clear from their paper that the
geometry BH found in the regions exterior to the active region is
exactly the same as that found by Ernst \cite{Ernst} for the
electromagnetic geon.}  This means that the details of the active
region as discussed by Wheeler \cite{Wheeler} are identical
(mutatis mutandis) to those of the gravitational geon found by BH. 
There is of course
a difference in the wave equations.  However, the effective time
averaged stress-energy of the gravitational waves is, to leading
order, identical to that of the electromagnetic waves.  

     In Sec.\ II a review of the BH solution is given.  In
Sec.\  III a self-consistent expansion of the wave equation and
the effective stress-energy tensor for the gravitational waves is
presented.  Explicit leading order equations are derived in Sec.\ 
IV, and it is shown that in the active region the equations to
leading order are those found by Wheeler for the electromagnetic
geon.  Extension of the calculation beyond the leading order is
also discussed.  Our conclusions are summarized in Sec.\ V.

\section{The Brill-Hartle Solution}

     The gravitational geon found by BH is a solution to Einstein's
equations that consists of gravitational waves propagating on a
static spherically symmetric background, which is created by the
waves.  There is a
large number of waves, each with a small amplitude, a high
frequency, and a large angular momentum.   
The waves have different angular orientations and somewhat
different frequencies.  The stress-energy
of the waves is significant in and near a spherical shell called
the active region, and insignificant elsewhere.  The solution is
self-consistent in that the geometry produced by the stress-energy
of the waves traps them in the active region for a long time.

     We begin by reviewing the BH gravitational wave expansion and
the BH definition of an effective stress-energy tensor for the
gravitational waves.  Consider a separation of the metric into a
background part, $\gamma_{\mu \nu}$, and a perturbation, $h_{\mu
\nu}$,      
\begin{equation}
    g_{\mu \nu} = \gamma_{\mu \nu} + h_{\mu \nu}.
\end{equation}
The Einstein tensor can also be divided into a part describing the
curvature due to the background geometry and that due to the
perturbation,
\begin{equation}
  G_{\mu \nu}(g_{\alpha \beta}) = G_{\mu \nu}(\gamma_{\alpha
\beta}) + \bigtriangleup G_{\mu \nu}(\gamma_{\alpha
\beta},h_{\alpha \beta})  \;\;\;.
\end{equation}
It is important to note that $\bigtriangleup G_{\mu \nu}$ is
defined by this equation.  

One way to specify the separation (1) is to use some smoothing or
averaging procedure $< \,>$ acting on $G_{\mu \nu}$, and demand
that this procedure
not affect the value of $G_{\mu \nu}(\gamma_{\alpha \beta})$. 
In addition, one demands both the exact
and the averaged Einstein equations for vacuum. This leads to
\begin{mathletters}
\begin{eqnarray}
\bigtriangleup G_{\mu \nu}(\gamma,h) &=& <\bigtriangleup G_{\mu
\nu}(\gamma,h)> \\
G_{\mu \nu}(\gamma) &=& - <\bigtriangleup G_{\mu \nu}(\gamma,h)> 
\;\;\;.
\end{eqnarray}
\end{mathletters}%
(Here and hereafter the indices of the arguments of $G_{\mu \nu}$
and $\bigtriangleup G_{\mu \nu}$ are suppressed for notational
simplicity.)
The first equation is regarded as a wave equation, which gives
the behavior of the perturbed geometry.  The second equation is the
backreaction equation, which describes how the background geometry
is affected by the perturbed geometry.  From the
backreaction equation it is natural to define an effective 
stress-energy tensor for the gravitational waves as
\begin{equation}
<T_{\mu \nu}> = - {1 \over {8 \pi}} <\bigtriangleup G_{\mu
\nu}(\gamma,h)>  \;\;\;.
\end{equation}

For any valid perturbation expansion $\bigtriangleup
G_{\mu\nu}$ can be written in the form
\begin{equation}
 \bigtriangleup G_{\mu\nu} =  \bigtriangleup_1 G_{\mu \nu} +
\bigtriangleup_2 G_{\mu \nu} + ... \;\;\;.
\end{equation}
BH obtained an expansion of the form (5) by first expanding
$\bigtriangleup G$ in powers of $h$ and its derivatives.  They then
considered the high frequency, large angular momentum limit of this
expansion.  Thus their leading order term $\bigtriangleup_1 G$
consists of the high frequency large angular momentum limit of the
terms in the original expansion which are linear in $h$.  Their
second order term $\bigtriangleup_2 G$ consists of the appropriate
high frequency, large angular momentum limits of the terms in the
original expansion which are quadratic in $h$.\footnote{It also contains
certain terms from the original expansion that are linear in $h$.  However 
these vanish when a time average is taken.}
BH used
a time averaging for their stress-energy tensor.  This plus the
high frequency of the waves resulted in a stress-energy tensor
consisting of the time average of the high frequency large angular
momentum limits of the quadratic terms in the original expansion of
$\bigtriangleup G$ in powers of $h$ and its derivatives. 
To completely fix the value of the stress-energy tensor 
they implicitly made the choice\footnote{It is shown in 
Ref.(\cite{stressten}) that Eq.\ (4) does not completely specify the
definition of the stress-energy tensor.  An extra condition must be
imposed to determine uniquely the value of $<\bigtriangleup G>$.  
Although not presented as such in their paper, the condition
$<\bigtriangleup_1 G> = 0$ imposed by BH served to fix the value
of $<\bigtriangleup G>$ in the gauge they worked in.  See Sec.\ 
III.A for further discussion.}
$<\bigtriangleup_1 G_{\mu \nu}> = 0$. 
The resulting wave and backreaction equations are
\begin{mathletters}
\begin{eqnarray}
& & \mbox{} \triangle_1 G_{\mu \nu}(\gamma,h) = 0  \\
& & \mbox{} G_{\mu \nu}(\gamma) = -<\bigtriangleup_2 G_{\mu
\nu}(\gamma,h)> \;\;\;.
\end{eqnarray}
\end{mathletters}%

     The background metric for the geon solution is static and
spherically symmetric, so it can be written in the form
\begin{equation}
  \gamma_{\mu \nu} = $diag$(-e^\nu, e^\lambda, r^2, r^2 \sin^2
\theta) \;.
\end{equation}
BH used a variational approach to find their solution, the essence
of which is to show that the effective averaged stress-energy
tensor has vanishing trace to leading order. 
This suffices to define the metric outside the active region, 
which is near $r = a$.  The value of $a$ depends on the
ratio of the angular momentum to the frequency of the waves.  In
the infinite frequency and angular momentum limits the active
region shrinks to an infinitely thin shell at $r=a$. 
BH's self-consistent treatment found in this limit
that 
\begin{eqnarray}
e^\nu &=& {1 \over 9}, \;\; e^{-\lambda} = 1 \qquad\;\;\; r < a \;,
\nonumber \\
e^\nu &=&  e^ {-\lambda} = 1 - {{2 M}\over r} \quad\;\;    r > a
\;,
\end{eqnarray}
and that the mass of the geon is $M = (4/9) a$.  
Examination of Eq.\ (8)
suggests that, in the active region, $\partial \lambda / \partial
r$ is large while $\partial \nu / \partial r$ is of order
unity.\footnote{We
use the term ``unity" to mean the power $a^n$ of the geon radius
$a$, with $n$ adjusted so ``unity" has the same dimension as the
quantity with which it is being compared.}

\section{Consistency of the Gravitational Wave Expansion}

Several consistency questions must be answered to validate the BH
solution to Eqs.\ (6a) and (6b).
First, the effective stress-energy tensor of the gravitational
waves, Eq.\ (4), 
must be conserved with respect to the background geometry $\gamma$
in order 
to be a source of this background geometry. Second, the solution
ansatz for the wave equation must be sufficiently general so that
this set of equations can be simultaneously solved. 
Third it is necessary to investigate gauge
invariance.  For the wave and backreaction equations to be
consistent, the effective stress-energy tensor must be at least
approximately gauge invariant if the wave equation is.  Also  BH
used a gauge transformation to reduce the number of components of
$h_{\mu\nu}$.
Finally it is necessary to show that the expansion of
$\bigtriangleup G$ used by BH is a valid one for the geon
solution.

It is appropriate to note that we are here constructing one
particular type of geon, as described in the beginning of Sec.\ 
II. No attempt is made to discuss other possible gravitational
geons, such as thick-shell ones \cite{power}. To show existence of
geon solutions 
it is only necessary to establish that one particular,
carefully selected set of waves does form a geon.

\subsection{Conservation and gauge invariance}

Conservation of the effective stress-energy tensor for
gravitational waves can be established in a straight-forward
manner.  If we have a solution of Eq.\ (3b), then the Bianchi
identity satisfied by 
$G_{\mu \nu}(\gamma)$ implies that the exact stress-energy
tensor (4) is conserved with respect to the background
geometry $\gamma_{\mu\nu}$.  Thus the approximate stress-energy
tensor is similarly conserved to leading order so long as a 
self-consistent expansion is used for the gravitational waves.

     The question of whether the wave equation and the
effective stress-energy tensor are gauge invariant is much more
difficult to answer.  It is useful in this regard to discuss some
general results, first for the wave equation and then for the
stress-energy tensor.

     If the background geometry $\gamma$ is a solution to the
vacuum Einstein equations then it can be shown that if an expansion
of the form (5) is used with $\bigtriangleup_n G$ consisting of
terms of $n$th order in $h$ and its derivatives, then the wave
equation (6a) is exactly invariant under gauge transformations of
the form
\begin{equation}
\bar{h}_{\mu\nu}(x) = h_{\mu\nu}(x) - \gamma_{\mu \alpha}(x)
{\xi^\alpha}_{,\nu} - \gamma_{\alpha \nu}(x) {\xi^\alpha}_{,\mu}
- \gamma_{\mu\nu,\alpha} \xi^\alpha \;\;\;.
\end{equation}
Here $\xi$ results from an arbitrary coordinate transformation of
the form
\begin{equation}
\bar{x}^\mu = x^\mu + \xi^\mu \;\;\;.
\end{equation}
 If the same type of expansion is used but the background geometry
is a solution of Eq.\ (6b) and if the transformations are restricted
so that $\xi$ and its derivatives are no larger in magnitude than
$h$ and its derivatives, then it can be shown that the wave
equation is invariant to $O(h^2)$ under the above type of gauge
transformation.  For high frequency, large momentum gravitational
waves and the same restrictions on $\xi$, Isaacson\cite{I} showed
that the leading order wave equation is invariant under gauge
transformations of the form (9) to second order in his expansion. 
It is not hard, using an argument similar to that used by Isaacson,
to show that the wave equation for the gravitational geon solution
is also invariant to second order under gauge transformations of
the form (9) so long as $\xi$ and its derivatives are no larger in
magnitude than $h$ and its derivatives.  Here and hereafter by $n$th
order we mean of the same order as $\bigtriangleup_n G(\gamma,h)$ in
a self-consistent expansion of the form (5).

     Gauge invariance of the effective stress-energy tensor is a
different matter.  For Eqs.\ (6a) and (6b) to be consistent, the
stress-energy tensor must at least be gauge invariant to second
order.  However, Isaacson showed that $\bigtriangleup_2 G$ is not
invariant under transformations of the form (9).  The only case he
found in which the stress-energy tensor and thus the backreaction
equation (6b) are gauge invariant to second order is the case of
high frequency, large momentum waves when the averaging is over a
region of spacetime which is large compared to the wavelengths of
the waves but small compared to the scale on which the background
geometry varies.

     One might hope that in the geon case, where high frequency
large angular momentum waves are used, that the stress-energy
tensor would similarly be gauge invariant to second order.  However,
an explicit calculation using the gauge transformation (17) below
shows that this is not the case.  In fact, even when the background
geometry is a solution to the vacuum Einstein equations (including
the case of the flat space solution), the stress-energy tensor is
not gauge invariant if time averaging rather than spacetime
averaging is used.  Thus it appears that Eqs.\ (6a) and (6b), which were
implicitly solved by BH to obtain the geon solution, are
inconsistent when time averaging is used.

     The resolution to this very serious problem is given in
Ref.\cite{stressten}.  It is as follows:  First Eqs.\ (6a) and (6b) must be
replaced by the equations that result from substituting Eq.\ (5) into
Eqs.\ (3a) and (3b).  The result to $n$th order is
\begin{mathletters}
\begin{eqnarray}
\bigtriangleup_1 G + ... + \bigtriangleup_n G &=&
<\bigtriangleup_1 G + ... + \bigtriangleup_n G> \\  
G(\gamma) &=& -<\bigtriangleup_1 G + ... + \bigtriangleup_n G> 
\;\;\;.
\end{eqnarray}
\end{mathletters}%
Then a generalized gauge transformation is used.
It is arrived at by using the coordinate transformation (10) and
not allowing the functional form of the background geometry to
change under this coordinate transformation, that is
${\bar{g}}_{\mu\nu} = \gamma_{\mu\nu} + {\bar{h}}_{\mu\nu}$. Then 
${\bar{h}}_{\mu\nu}$ is given implicitly by the equation
\begin{eqnarray}
\gamma_{\mu \nu}(x) + h_{\mu \nu}(x) &=& \gamma_{\mu
\nu}(\bar{x}) + \bar{h}_{\mu \nu}(\bar{x}) + (\gamma_{\mu
\alpha}(\bar{x}) + \bar{h}_{\mu \alpha}(\bar{x}))
{\xi^\alpha}_{,\nu} \nonumber \\
&   &  + (\gamma_{\alpha \nu}(\bar{x}) + \bar{h}_{\alpha
\nu}(\bar{x})) {\xi^\alpha}_{,\mu} + (\gamma_{\alpha
\beta}(\bar{x}) + \bar{h}_{\alpha \beta}(\bar{x}))
{\xi^\alpha}_{,\mu} {\xi^\beta}_{,\nu}
\end{eqnarray}
Here derivatives of $\xi$ are with respect to $x$ not $\bar{x}$. 
The generalized gauge transformation is defined as one in which the
quantity $\bar{h}(x)$ is substituted for $h(x)$ into the
expression of interest.  If $h$, $\xi$ and their derivatives are
small enough, then to leading order this gauge transformation
reduces to the usual transformation (9).  

It is proven in Ref.\cite{stressten} that, when the vacuum field
equations are satisfied, $\bigtriangleup G_{\mu\nu}$ is invariant
under this generalized gauge transformation.  
It is also argued in that paper that the 
quantity
\begin{displaymath}
  \bigtriangleup_1 G_{\mu\nu} + ... + \bigtriangleup_n G_{\mu\nu}
\end{displaymath}
is gauge invariant to $n$th order.  This implies that
Eqs.\ (11a) and (11b) are invariant to $n$th order under generalized gauge
transformations.  Thus there is no problem
with gauge invariance to any order so long as generalized gauge
transformations are used.
     
     In the same paper it was shown that there is a large amount of
freedom available in choosing the form of the effective
stress-energy tensor for gravitational waves.  This freedom is
related to the freedom one has in choosing 
in the split between the background and the perturbed geometry. 
Thus, as noted above, the BH ansatz
\begin{equation}
<\bigtriangleup_1 G_{\mu\nu}> = 0
\end{equation}
is simultaneously a choice of the form of the stress-energy
tensor and a definition of the
separation between the background and perturbed geometry that is to
be used.  From the above discussion 
it is clear that the condition (13) is, in general, only gauge
invariant to first order.  The lack of exact gauge invariance in
this condition is a reflection of the fact that the splitting
between the background and the perturbed metric is inherently gauge
dependent.  

As mentioned above, an explicit calculation 
which we have made
using the gauge
transformation (17) below shows that, even 
after time averaging,
$<\bigtriangleup_2 G_{\mu\nu}>$ 
is not separately gauge invariant.  The problem is simply
that the condition (13) was used along with the
expansion (5) in the derivation of Eqs.\ (6a) and (6b).  However, this
condition
can only be imposed in a particular gauge since it is not exactly
gauge invariant.
Since BH solved Eqs.\ (6a) and (6b) and since we also use those
equations as the starting point of the derivations in Sec.\ IV, it
is necessary to show that these equations are 
consistent with the second order version of Eqs.\ (11a) and (11b).

To begin note that to actually solve Eqs.\ (11a) and (11b) in a particular
gauge one can
follow Isaacson\cite{I} and make the following expansion for $h$
\begin{equation}
  h = h^{(1)} + h^{(2)} + ... \;\;\;,
\end{equation}
where $h^{(n)}$ is defined such that $\bigtriangleup_1
G(\gamma,h^{(n)})$ is of 
the same order as
$\bigtriangleup_n G(\gamma,h^{(1)})$.  Then to
second order Eqs.\ (11a) and (11b) can be written
\begin{mathletters}
\begin{eqnarray}
\bigtriangleup_1 G(\gamma,h^{(1)}) &=& <\bigtriangleup_1
G(\gamma,h^{(1)})> \;\;\;, \\
\bigtriangleup_1 G(\gamma,h^{(2)}) + \bigtriangleup_2
G(\gamma,h^{(1)}) &=& <\bigtriangleup_1 G(\gamma,h^{(2)}) +
\bigtriangleup_2 G(\gamma,h^{(1)})> \\
G(\gamma) &=& - <\bigtriangleup_1 G(\gamma,h^{(1)}) +
\bigtriangleup_1 G(\gamma,h^{(2)}) + \bigtriangleup_2
G(\gamma,h^{(1)})> 
\end{eqnarray}
\end{mathletters}%
Condition (13) can be imposed by requiring that
$<\bigtriangleup_1
G(\gamma,h^{(n)})> = 0$ for all $n$.  Then to lowest order
Eqs.\ (15a)-(15c) are
equivalent to Eqs.\ (6a) and (6b) with the substitution $h \rightarrow
h^{(1)}$.

     It is important to emphasize that it is the original equations
(11a) and (11b)
that are gauge invariant to $n$th order and that 
Eqs.\ (15a) - (15c) and thus Eqs.\ (6a) and (6b) are in
general only correct in a particular gauge.  The procedure being
followed
here is the usual one for finding solutions to Einstein's
equations.  First
the equations are written down in a particular coordinate system
and
then they are solved (here in an approximate manner) 
in that coordinate system. 

\subsection{A valid expansion of the Einstein tensor}

     Since the gravitational waves making up the BH geon are high
frequency, large angular momentum waves, it is necessary to find an
expansion of the Einstein tensor that is appropriate for these
waves.  The key point for a thin-shell geon is, that while the 
waves' amplitudes are much
smaller than unity, their derivatives are large compared to unity. 
Further, while the background metric is of order unity, some,
but not all, of its derivatives are much larger than unity in the
active region (although they are of order unity or smaller well
outside of the active region). Thus, whether a term is of leading
order will be determined not only by the power of $h$ but also by
the power of the frequency $\omega$ or of the harmonic order \ls\ 
that it contains.

     What must be done is to find a self-consistent expansion of
the Einstein tensor with the above constraints on the background
metric and the perturbations.  Since some derivatives of the
background metric are large in the active region but of order unity
outside of it, this is a complicated task.  It clearly depends on
the solutions to both the wave equation and backreaction equation
which in turn depend on the expansion used.  

     Fortunately, as is shown by direct calculation in Sec.\ 
IV, to second order the method used by BH and discussed in Sec.\ 
II works.  However for the reasons discussed above, it would
be significantly more difficult to compute the third and higher
order terms in the correct expansion of $\bigtriangleup G$. 

\section{The wave and backreaction equations}

Derivations of the wave and backreaction equations have been attempted 
at the time of the BH paper \cite{Hartle} and subsequently by us
and independently by \cite{C2}.  Our version provides a correct, independent
treatment of these important equations.
We begin with the wave equation.

\subsection{The wave equation}

     The geon is composed of
a large number of gravitational waves and the equations for each
are the same.  BH used the equations $\bigtriangleup_1 R_{\mu \nu}
= 0$, with $R_{\mu \nu}$ the Ricci tensor.  BH showed that to leading order
these equations are equivalent  to (6a).  We first expand $\bigtriangleup
R_{\mu\nu}$ in powers of $h_{\mu\nu}$ and its derivatives, so the
wave equation is the high frequency and large angular momentum
limit of
\begin{equation}
 \gamma^{\alpha \beta} [h_{\mu \nu;\alpha \beta} + h_{\alpha \beta;
\mu \nu} - h_{\mu \alpha;\nu \beta} - h_{\nu \alpha;\mu \beta}] =
0 \;\;\;.
\end{equation}
Here indices as usual are raised and lowered by the background
metric $\gamma_{\alpha \beta}$.  

Following BH we analyze the angular behavior
of solutions to these equations for static spherically symmetric
spacetimes by the method of  
Regge and Wheeler \cite{RW}, and limit ourselves to
azimuthally symmetric, odd parity functions.  
In the gauge of Regge and Wheeler 
the perturbation $h$ is
obtained from the general $h$ via a gauge transformation of the
form
\begin{eqnarray}
  \xi^t &=& \xi^r = 0 \nonumber \\
  \xi^\mu &=& e^{-i \omega t} \Lambda(r) \epsilon^{\mu \nu}
\frac{\partial}{\partial x^\nu} {Y_l}^0 (\theta),  \;\;\;\;
(\mu,\nu = \theta, \phi)    \;\;\;.
\end{eqnarray}
Here $\Lambda(r)$ is some function of $r$, $\epsilon^{\mu\nu}$ is
an antisymmetric tensor with
$\epsilon^{3 2} = (r^2 \sin \theta)^{-1}$, and ${Y_l}^0 (\theta)$
is a spherical harmonic.  In their gauge the ``odd" solutions are
then written in the form
\begin{eqnarray}
  h_{\mu \nu} &=& r_{\mu \nu}e^{-i \omega t} \sin \theta \; {{d
{Y_l}^0 (\theta)} \over {d \theta}}  + c.c.\;,  \nonumber \\
  r_{\mu \nu}(r) &=& h_0(r) [{\delta_\mu}^0 {\delta_\nu}^3 +
{\delta_\nu}^0 {\delta_\mu}^3] + h_1(r)[{\delta_\mu}^1
{\delta_\nu}^3 + {\delta_\nu}^1 {\delta_\mu}^3] \;\;.
\end{eqnarray}  

     When these equations are substituted into Eq.\ (16), and when
the background metric is of the form (7), there result the
following three equations for $h_0$ and $h_1$:
\begin{mathletters}
\begin{eqnarray}
  i \omega e^{-\nu} \left({{d h_0} \over {d r}} - {{2 h_0} \over r}
\right) + h_1 \left[ {{{l^*}^2}\over {r^2}} - {{\omega}^2} e^{-\nu}
+ {{e^{-\lambda}}\over r} \left(\lambda_r - \nu_r - {2 \over r}
\right) \right] &=& 0     \\
  i \omega h_0 e^{-\nu} + e^{-\lambda} \left[ {1 \over 2} (\nu_r -
\lambda_r) h_1 + {{d h_1} \over {dr}} \right] &=& 0  \\
  {{d^2 h_0} \over{d r^2}} + i \omega \left[{{d h_1} \over {d r}}
+ h_1 \left( {2 \over r} - {1 \over 2} (\lambda_r + \nu_r) \right)
\right]
  - {1 \over 2} (\lambda_r + \nu_r) {{d h_0} \over {d r}} - \left[
{{e^\lambda {l^*}^2}\over {r^2}} - {{2 \nu_r} \over r} \right] h_0
&=& 0  \;.  
\end{eqnarray}
\end{mathletters}%
Here we have put ${l^*}^2 \equiv l(l+1)$, and denoted radial
derivatives 
of the background metric by subscripts.  From Eq.\ (19b) it is
clear that if we take $h_1$ to be real then $h_0$ is imaginary. 
Taking note of the complex conjugate term in Eq.\ (18) it is then
seen that the part of the disturbance associated with $h_0$ has a
time dependence of the form $\sin(\omega t)$ while the part of the
disturbance associated with $h_1$ has a time dependence of the form
$\cos(\omega t)$.

By using Eq.\ (19b) the function $h_0$ can be eliminated from
Eqs.\ (19a) and (19c).  Then the change of variables
\begin{mathletters}
\begin{eqnarray}
 d r^* &=& e^{{1 \over 2}(\lambda - \nu)} d r \;\;\;, \\
 Q &=& e^{{1 \over 2}(\nu - \lambda)} {{h_1} \over r} 
\end{eqnarray}
\end{mathletters}%
results in the two equations\footnote{ By taking the derivative of
(21a) with respect to $r^*$ it is
easy to show that the two equations are not consistent unless the
background metric $\gamma_{\mu \nu}$ is an exact solution to the
vacuum Einstein equations.  This is because for a general
background the lack of exact gauge invariance does not allow
the simplifying gauge transformation (17), as explained above.  
However, the approximate gauge invariance of the wave equation
results in the equations being 
consistent to leading order in \ls\ if the background 
geometry satisfies the backreaction equations to leading order.}
\begin{mathletters}
\begin{eqnarray}
\frac{d^2 Q}{d {r^*}^2} &+& 
 \left[{{\omega}^2} - {e^{\nu }}\,{{{{{l^*}^2}}}\over {{r^2}}} - 
   {e^{{{-\lambda }\over 2} + {{\nu }\over
2}}}\,{{{3\,\lambda_{r^*}}}\over 
     {2\,r}} + {e^{{{-\lambda }\over 2} + {{\nu }\over 2}}}\,
       {{{3\,\nu_{r^*}}}\over {2\,r}} \right] Q = 0 \\
\frac{d^3 Q}{d {r^*}^3} &+&
\left[-{\nu_{r^*}} + {e^{{{-\lambda }\over 2} + {{\nu }\over
2}}}\,{3\over r} \right] {{d^2 Q}\over{d {r^*}^2}}  
 + \left[ {{{\omega}}^2} - {e^{\nu }}\,{{{{l^*}^2}}\over {{r^2}}}
- {e^{{{-\lambda }\over 2} + {{\nu }\over
2}}}\,{{{3\,\lambda_{r^*}}}\over{2\,r}} + {e^{{{-\lambda }\over 2}
+ {{\nu }\over 2}}}\,{{{3\,\nu_{r^*}}}\over {2\,r}} \right]
{{dQ}\over{d {r^*}}} \nonumber \\
  &+& \left[ -{{{{\omega}}^2}\,\nu_{r^*}}  - 
   {e^{{{-\lambda }\over 2} + {{3\,\nu }\over
2}}}\,{{{{l^*}^2}}\over{{r^3}}} + {e^{-\lambda  +
\nu}}\,{{{2\,\nu_{r^*}}}\over {{r^2}}} + {e^{{{-\lambda }\over 2}
+ {{\nu }\over 2}}}\,{{{{{\lambda_{r^*}}}^2}}\over{4\,r}} \right.
\nonumber \\ 
& & \left. -{e^{{{-\lambda }\over 2} + {{\nu }\over
2}}}\,{{{\lambda_{{r^*} {r^*}}}}\over {2\,r}} - {e^{{{-\lambda
}\over 2} + {{\nu }\over 2}}}\,{{{{\nu_{r^*}}^2}}\over{4\,r}} +
{e^{{{-\lambda }\over 2} + {{\nu }\over 2}}}\,{{{\nu_{{r^*}
{r^*}}}}\over {2\,r}} + {e^{{{-\lambda }\over 2} + {{\nu }\over
2}}}\,{{{3\,{{\omega}}^2}}\over r} \right] Q = 0 \;\;.
\end{eqnarray}
\end{mathletters}%
To ascertain the correct limiting form of these equations we take
a hint from the EM geon's background  and put
\begin{eqnarray}
   \omega &\sim & l^* \nonumber \\
  {{dQ}\over {d{r^*}}} &\sim & l^* Q \nonumber \\
  {{d\lambda}\over {d{r^*}}} &<<& O({l^*})  \\
  {{d^2\nu}\over {{d{r^*}}^2}} &<<& O({l^*}) \;\;\;. \nonumber
\end{eqnarray}
(In the following subsections we  
show that this indeed allows self-consistent solutions.)
To leading order (\ls$^2$), 
Eqs.\ (21a) and (21b) are 
\begin{mathletters}
\begin{eqnarray}
{{d^2 Q} \over {d {r^*}^2}} + 
 \left[ {{{\omega}}^2} - {{{e^{\nu }}\,{{l^*}^2}}\over {{r^2}}}
\right] Q &=& 0 \\
\frac{d^3 Q}{d {r^*}^3} + \left[\omega^2 - \frac{e^{\nu}
{l^*}^2}{r^2} \right] \frac{d Q}{d {r^*}} &=& 0 \;\;. 
\end{eqnarray}
\end{mathletters}

Note that these equations are consistent to leading order in
$l^*$.  Note also that it is not being assumed here that
$d\lambda/d{r^*}$ and $d^2 \nu/{d{r^*}}^2$ are of order unity or
less, simply that they
are not of order $l^*$.  It turns out that for the geon solution
both of these quantities are of order unity outside of the active
region, but they are of order ${l^*}^{2/3}$ inside the active
region.  It also turns out that $dQ/d{r^*}$ is of order ${l^*}\, Q$
outside the active region, but is of order ${l^*}^{2/3}\,Q$ inside
the active region.  Thus Eqs.\ (23a) and (23b) describe the leading order
behavior of the gravitational waves, both inside and outside of the
active region.

\subsection{The backreaction equations}

     The next step in deriving the equations that result in a
gravitational geon solution is to consider the effective 
stress-energy tensor, Eq.\ (4), for the gravitational waves.  As with
the wave
equation, the easiest way to derive the stress-energy tensor to
leading order is to consider the linear and quadratic terms in an
expansion of the Einstein tensor in powers of $h_{\mu \nu}$.  The
large \ls \ limit of these terms is to be taken in the same way as
it was for the wave equation in the previous subsection.  The
result is averaged over times which are long compared with the
gravitational wave period.

The background metric for the geon solution is static and
spherically symmetric, so from the backreaction equation (6b) it is
clear that the stress-energy tensor should have these same
symmetries.  
Time averaging automatically makes it static over the geon lifetime. 
Spherical symmetry is a more difficult problem. 
Arranging the waves so that the stress-energy tensor
is spherically symmetric has been discussed in detail by Wheeler
\cite{Wheeler} in the case of the electromagnetic geon.  The
arrangements are identical for the gravitational geon.  They are
briefly summarized here.

     First, as can be seen from Eq.\ (18), to have spherical symmetry
in the large \ls \, limit it is necessary to have many
gravitational waves with different angular orientations. Thus we
put 
\begin{equation}
  h_{\mu \nu} = \sum_i (h_{\mu \nu})_i 
\end{equation}
with each $(h_{\mu \nu})_i$ a solution to Eq.\ (6a) with a different
value of $\omega$, a different phase, and a different polar
axis.  If the phases are random and the values of $\omega$,
while different, are all approximately equal and large, then after
the time averaging is done, to leading order in $\omega$ the cross
terms in $<\bigtriangleup_2 G_{\mu \nu}>$ vanish and  
\begin{equation}
<\bigtriangleup_2 G_{\mu \nu}>\, \approx \sum_{i}
\,<\left(\bigtriangleup_2 G_{\mu \nu} \right)_i> \;.
\end{equation}

     It is useful to denote
$(\bigtriangleup_2 G_{\mu \nu})_I$ as the wave which has a pole at
$\theta = 0$.  Wheeler \cite{Wheeler} has shown that if the
distribution of the polar axes of the waves is uniform, then when
the waves are all added together one has
\begin{eqnarray}
     <\bigtriangleup_2 {G_t}^t> &=& \frac{1}{2} \int_{0}^{\pi}
<\left(\bigtriangleup_2 {G_t}^t\right)_I> \sin \theta \,d \theta
\nonumber \\
     <\bigtriangleup_2{G_r}^r> &=& \frac{1}{2} \int_{0}^{\pi}
<\left(\bigtriangleup_2 {G_r}^r\right)_I> \sin \theta \,d \theta \\
     <\bigtriangleup_2{G_\theta}^\theta> &=& \frac{1}{4}
\int_{0}^{\pi}
\left(<\left(\bigtriangleup_2{G_{\theta}}^{\theta}\right)_I> +
<\left(\bigtriangleup_2 {G_{\phi}}^{\phi}\right)_I>\right) \sin
\theta \, d \theta \;\;. \nonumber  
\end{eqnarray}

     Finally there is the question of conservation.  From the
discussion of Sec.\  III.A it is clear that $<T_{\mu \nu}>$ should
be automatically conserved to leading order.  Since the geon is
here treated 
to leading order, that is in the large \ls\ limit, one expects
conservation to 
hold.  Direct
calculation shows that the stress-energy tensor is conserved to
leading order in this limit.

     In the large \ls \ limit the backreaction equations coming
from the $(t,t)$, $(r,r)$ and $(\theta,\theta)$ components of (6b)
in an orthonormal frame are
\begin{mathletters}
\begin{eqnarray}
  -{1\over {{r^2}}} + 
   {{e^{-{\lambda}}}}\,{1\over {{r^2}}} - 
   {e^{{{-{\lambda}}\over 2} - {{\nu}\over 2}}}\,
       {{{\lambda}_{r^*}}\over r}
   &=& \frac{-{l^*}^4}{8\,\left( 2\,l + 1 \right)
\,{{{\omega}}^2}\,{r^4}}
 \left[ {l^*}^2\,
        {{Q}^2} + e^{-\nu}{{{\omega}}^2}\,{r^2}\,{{Q}^2} + 
     e^{-\nu}{r^2}\,{\left(\frac{dQ}{d{r^*}}\right)^2} \right]   \\
 -\frac{1}{r^2} + {{e^{{-\lambda}}}}{1\over {{r^2}}}+ 
  {{e^{{{{-\lambda}}\over 2} - {{\nu}\over 2}}}\,
       {\nu_{r^*}}\over r} 
   &=& \frac{{l^*}^4}{8\,\left( 2\,l + 1 \right)
\,{{{\omega}}^2}\,{r^4}}
 \left[-{{{l^*}}^2}\,{{Q}^2} + e^{-\nu}{\omega}^2\,{r^2}\,{{Q}^2}
\right. \nonumber \\
 & & \left. -e^{-\frac{\lambda}{2} - \frac{\nu}{2}}\,6\,
r\,Q\,{\frac{ dQ}{d{r^*}}}  + e^{\frac{-\lambda}{2} -\frac{3\nu}{2}}
       \frac{8\,{{{\omega}}^2}\,{r^3}}{{l^*}^2}\,Q\,{\frac{
dQ}{d{r^*}}} + e^{-\nu}{r^2}\,{{{\left(\frac{ dQ}{d{r^*}}\right)}}^2} \right]  \\
 e^{{-(\lambda + \nu)}\over 2}  \,\left(
       {{\nu_{r^*}-{\lambda}_{r^*}}\over {2 r}}\right) + 
   e^{-\nu}{{\,\nu_{{r^*}{r^*}}}\over {2}}
&=& \frac{{l^*}^4}{8\,{\left( 2\,l +1 \right)
\,{{{\omega}}^2}\,{r^4}}}
  \left[-2 \,{{{{l^*}}}^2}\,
      {{Q}^2} + e^{-\nu} 7\,{{{\omega}}^2}\,{r^2}\,{{Q}^2} \right.
\nonumber \\
& & \left. - e^{-2\nu}\,
\frac{4\,{{{\omega}}^4}\,{r^4}}{{l^*}^2}\,{{Q}^2}     -
e^{-\nu}3\,{r^2}\,{\left({{\frac{ dQ}{d{r^*}}}}\right)^2}    + {e^{-2\nu}}
\frac{4\,{{{\omega}}^2}\,{r^4}}{{l^*}^2}\,{\left({{\frac{
dQ}{d{r^*}}}}\right)^2} \right] 
\end{eqnarray}
\end{mathletters}%
In arriving at these equations the following identities were used
\begin{eqnarray}
\frac{1}{2} \int_{-1}^1 (P_l(x))^2 dx &=& \frac{1}{2 l + 1} \nonumber \\
\frac{1}{2} \int_{-1}^1 (x P_l(x) {P_l}'(x) dx &=& \frac{l}{2 l + 1} \nonumber \\
\frac{1}{2} \int_{-1}^1 ({P_l}'(x))^2 dx &=& \frac{l(l+1)}{2} \\
\frac{1}{2} \int_{-1}^1 (1 - x^2) ({P_l}'(x))^2 dx &=& \frac{l(l+1)}{(2 l + 1)} \;.
     \nonumber
\end{eqnarray}
Here $P_l(x)$ is a Legendre Polynomial.

The combined set of Eqs.\ (23a), (23b), (27a), (27b), and (27c) completely
specify the background geometry and the gravitational waves in the
high frequency and angular momentum limit.  In the next subsection
Wheeler's expansion is used to show that in and near the active
region it is possible to arrive at a set of equations which are
independent of both $\omega$ and \ls.

\subsection{Wheeler's expansion for the active region}

     The active region is defined as the region in which the
gravitational waves undergo radial oscillations\footnote{By radial oscillations
we mean here that the waves oscillate as a function of the radius, not
that they oscillate in the radial direction.} rather than radial
damping, and where their amplitudes are significant.  As can be seen
from Eq.\ (23a), the waves will radially oscillate whenever $\omega^2
> e^{\nu} {l^*}^2/r^2$.  This happens at large $r$, but in order to
construct a geon that lasts for a long time, we must use waves
whose amplitudes are completely negligible there.  Radial oscillation
also happens near $r=a$, where from Eqs.\ (8) and (23a) one finds\footnote{Until 
the equations are solved, the location of the active region
is unknown.  Wheeler\cite{Wheeler} initially expanded about $\rho
= l^*$.  After solving the equations he found the active region was near 
$\rho = l^*/3$.  He used a
scale invariance present in the original equations for the EM geon
(and also present in Eqs.\ (23a) and (23b) and (27a)-(27c)) to rescale the 
solutions accordingly.
To facilitate comparison with the equations derived by Wheeler we
also expand about $\rho = l^*$.}
\begin{equation}
a \sim \frac{l^*}{\omega}\;\;\;.
\end{equation} 

By expanding all of the relevant quantities in powers of
${{l^*}}^{1/3}$, Wheeler was able to arrive at a set of equations
which
describe the electromagnetic geon in and near the active region and
which do not explicitly depend on the value of \ls.  A similar
derivation is presented here.  Before making the expansions
it is useful first 
to change to new variables,
\begin{mathletters}
\begin{eqnarray}
   r &=& \frac{\rho}{\omega}  \\
   e^{-\lambda} &=& 1 - {{2 L(\rho)}\over \rho} \\
   e^{\nu} &=& \left(1 - {{2 L(\rho)}\over \rho}\right) S^2(\rho) \\
   Q &=& \frac{[8 (2 l + 1)]^{1/2}}{{l^*}^2} f(\rho) \;\;\;. 
  \end{eqnarray}
\end{mathletters}%
To remove the dependence on $l^*$ a new radial
variable $x$ is defined such that
\begin{mathletters}
\begin{eqnarray}
  x &=& (\rho^* - l^*) {{l^*}}^{-1/3} \\
  dx &=& {{l^*}}^{-1/3} d \rho^* \;\;\;.
\end{eqnarray}
\end{mathletters}%
Then the following expansions are made 
\begin{mathletters}
\begin{eqnarray}
  & \rho& = l^* + {l^*}^{1/3} {r_0}(x) + ... \;\;,  \\
  & L &= l^* \lambda_0 (x) + {l^*}^{2/3} \lambda_1 (x) +
{l^*}^{1/3} \lambda_2 (x) + ... \;\;,  \\
  & S &= \frac{1}{k(x)} + {l^*}^{-1/3} q_1 (x) + {l^*}^{-2/3} q_2
(x) + ... \;\;,  \\
  & f &= {l^*}^{1/3} \phi(x) + \phi_1 (x) + {l^*}^{-1/3} \phi_2 (x)
+ ... \;\;,  \\
& &  \left[1 - \left(\frac{l^* S}{\rho} \right)^2 \left(1 -
\frac{2 L}{\rho}\right) \right] = {l^*}^{-2/3} j(x) k(x) + ...
\;\; . 
\end{eqnarray}
\end{mathletters}%
It is next useful to 
derive several intermediate identities.  Using Eqs.\ (32a)-(32c) in
Eq.\ (32e) one finds
\begin{mathletters}
\begin{eqnarray}
\lambda_0 &=& \frac{1}{2} (1 - k^2) \\  
\lambda_1 &=& q_1 k^3
\end{eqnarray}
Combining Eqs.\ (20a), (30a), (31b), (32a), and (33a) one finds
\begin{equation}
\frac{d r_0}{d x} = k
\end{equation}
\end{mathletters}%
Finally Eqs.\ (23a), (27a), (27b), (30a)-(30d), (31a), (31b), (32a)-(32e), 
(33a)-(33c) can be combined to obtain, to leading order 
in \ls,  the equations
\begin{mathletters}
\begin{eqnarray}
  \frac{d^2 \phi}{d x^2} + j k \phi &=& 0 \;\;, \\
  \frac{d k}{d x} + \phi^2 &=& 0 \;\;, \\
  \frac{d j}{d x} - 3 + \frac{1}{k^2} \left[1 + \left(
\frac{d\phi}{d x} \right)^2 \right] &=& 0 \;\;.
\end{eqnarray}
\end{mathletters}

     These equations are exactly the same equations that Wheeler
\cite{Wheeler} found for the electromagnetic geon.\footnote{Cooperstock
et. al. \cite{C2} found a different set of equations.  Part of
the difference is related to the fact that they used a different normalization
for the variable $f$ in Eq.\ (30d).  This does not lead to qualitatively
different solutions.  However, sign errors in their equations do result
in qualitatively different solutions.  The right hand sides of their 
Eqs.\ (4.22) and (4.23) have the wrong signs given the sign conventions used in their 
paper.  These sign errors result in overall sign errors on the right hand
sides of their Eqs.\ (4.30), (4.31), and (4.46).  They also result in a sign error
in the last term on the right hand side of Eq.\ (4.47).} 
Thus we see
that, to leading order, the geometry of the gravitational geon is
exactly the same as that for the electromagnetic geon.  This was
already found to be the case by BH for the spacetime outside of the
active region.  It has now been shown to be true for the geometry
inside of the active region.  

     This result was hinted at when BH noted that in the large
frequency and angular momentum limit the equation of motion for the
gravitational waves is identical to that for electromagnetic waves.
It is also consistent with the result of Isaacson \cite{I} that
when the effective stress-energy tensor for high frequency, large
momentum gravitational waves is averaged over a small region of
spacetime, it is of the same form as that for electromagnetic
waves.

For a thin shell geon it is necessary that the solution to the wave equation 
(34a) fall off rapidly on both sides of the active region. Equations (34)
are therefore similar to an eigenvalue problem in which the amplitude of
$\phi$ (as measured, for example, by its peak value) plays the role of the 
eigenvalue: through equations (34b, c) the amplitude determines the strength
of the ``potential" $jk$, which must be just right to have $\phi$ as a
zero energy eigenfunction. For the case of the lowest eigenfunction these
equations have been discussed numerically by Wheeler \cite{Wheeler}, 
and more recently by Cooperstock et al \cite{C2}. The geon metric outside the active region, Eq.\ (8),
was first derived from the numerical solution.\footnote{Wheeler found $e^\nu 
\approx 0.11$, and Ernst \cite{Ernst} showed $e^\nu = 1/9$ exactly. The method of BH
yields the same exact result for all fields that form geons and whose  
(effective) stress-energy tensor has vanishing trace. Our development in the 
Appendix yields yet another way to derive the exact result.} 
Work is currently in
progress to investigate the properties of the other solutions.
The results will be presented elsewhere.

Although the
numerical results constitute strong evidence for the existence of geon
solutions, without an existence proof of exact solutions to the basic 
equations (34) the viability of any type of geon is open to 
doubt.\cite{C2} We provide the existence proof in the
Appendix.
     
     A few words are in order about extensions to higher orders in
the expansion. In the present approximation the waves are a kind of
``null
fluid" that can be isotropically distributed in the active region.
If the expansion of the Einstein
tensor is carried out to higher order in \ls\, the ``graininess" of
the stress-energy due to non-negligible cross terms of
finite wavelength in the
stress-energy tensor becomes apparent. This invalidates Eq.\ (25)
and makes the stress-energy tensor much more difficult to
compute.
The background geometry could however remain spherically
symmetric by a different choice of splitting between waves and
background.

\section{Conclusions}

     We have obtained a correct self-consistent set of equations
for the gravitational geon, which
describe the gravitational waves and the background geometry.  
These equations are
accurate in the high frequency, large angular momentum limit.  In
and near the active region they have been shown to be the same set
of equations as those found by Wheeler for the electromagnetic
geon.  Thus, to leading order, the geometry both inside and outside
of the active region of the gravitational geon is identical to that
of the electromagnetic geon.  

\acknowledgments
     P.R.A. would like to thank J. Hartle for suggesting this
problem and for some very helpful discussions.  This work was
supported in
part by Grant Number PHY-9512686 from the National Science
Foundation.

\appendix\section*{Discussion of the geon equation}

Our geons are taken to be governed by Eqs.\ (34). These equations
focus on the active region, and neglect the other oscillating region 
(beyond $r \sim 2.67 M$, where $\omega^2 > e^v {l^*}^2/r^2$, according 
to Eq.\ (8)). We can therefore solve Eq.\ (34a) as a true bound state problem
in the effective potential $jk$, rather than a virtual bound state. 
The appropriate boundary condition on $\phi(x)$ is therefore
\begin{equation} 
 \phi(-\infty) = \phi(+\infty) = 0 \; ,
\end{equation}
  From Eq.\ (34b) it is seen that we can follow Wheeler\cite{Wheeler} and
choose the boundary condition on $k$ to be
\begin{equation}
k(-\infty) = 1.
\end{equation}
(Since $e^\nu|_{r=0} = k^2|_{x=-\infty}$ this contradicts Eq.\ (8), and is due to  
the use of Wheeler's expansion before rescaling; see footnote 10. Nevertheless,
the ratio $e^\nu|_{r=0}/e^\nu|_{r=\infty} = k^2|_{x=-\infty}/k^2|_{x=\infty}$
will be correctly given.) A third condition follows from Eq.\ (34c),
\begin{equation}
j(-\infty) = -\infty.
\end{equation}
We cannot expect to solve Eqs.\ (34) with any more conditions than these three;
the integration of the equation itself will tell us whether solutions
of the geon type are possible, and what their remaining properties are.

We reduce the system (34) to a single equation by considering $k$ as the 
independent variable and using Eq.\ (34b) to convert derivatives, $d/dx =
-\phi^2 d/dk$. We will denote $d/dk$ by a prime ($'$). The unknown function 
$j(k)$ can then be eliminated between the two remaining equations. The
resulting third order equation is conveniently written in terms 
of a new function
\begin{equation}
H = F^2 F'' - 2 - 6 k^2 \qquad {\rm where} \qquad F = \phi^2,
\end{equation}
with the result
\begin{equation}
-H + kH' = 0.
\end{equation}
This can be integrated to yield $H = Ak$, with $A$ a constant of integration.
Thus the equation to be solved is
\begin{equation}
F^2 F'' = Ak + 2 + 6k^2
\end{equation}

To evaluate the constant $A$ we use the boundary conditions (A1) and (A2) at
$x = -\infty$. Because $F^2F'' = d^2F/dx^2 - F^{-1}(dF/dx)^2$, Eq.\ (A6) must yield zero at
$k = 1$, hence $A = -8$. Next we use (A1) at $x = +\infty$ to conclude that Eq.\ (A6)
also vanishes for $k(\infty)$, so that 
\begin{equation}
 k(\infty) = {1\over 3}.
\end{equation}
In view of Eqs.\ (30c), (32c) and the rescaling this yields the geon metric
(8) outside the active region (with $M$ not yet determined in terms of $\phi$).
It is appropriate to rewrite Eq.\ (A6) in terms of a shifted variable, $u = k -
{2\over 3}$ that exhibits the symmetry of this equation,
\begin{equation}
F^2 F'' = 6u^2 - {2\over 3}
\end{equation}
The boundary conditions (A1) and (34a) now show that $\phi$, $F$ and $jk$ 
are even functions of $u$, and by Eq.\ (34b) we can choose $x$ to be an odd 
function of $u$. Thus our four boundary conditions can be replaced by the
more convenient form
\begin{mathletters}
\begin{eqnarray}
F' = 0 \quad {\rm at} \quad u = 0 \\
F \rightarrow 0 \quad {\rm when} \quad u \rightarrow 1/3 \\ 
x = 0 \quad {\rm at} \quad u = 0 \\
2jF = 1 \quad {\rm at} \quad u = 0
\end{eqnarray}
\end{mathletters}
Of course only Eqs.\ (A9a) and (A9b) are needed to solve Eq.\ (A8).

To show existence of solutions it is enough to concentrate on the interval
$I : 0 \leq u \leq 1/3$. Starting with some positive initial value $F(0)$ and
condition (A9a), we can always integrate Eq.\ (A8) to larger and larger $u$
as long as $F(u)$ remains bounded away from zero. If $F(0)$ is sufficiently
large, the solution stays positive to $u = 1/3$. For example, the estimate
$F^2F'' =  6u^2 - {2\over 3} > -{2\over 3}$ shows that 
$F(0) > {1\over 3}\left({16\over \pi^2}\right)^{1/3} \approx 0.392$ is  
sufficient. We consider all solutions that are positive
in the interval $I$, and call them ``solutions in $I$" for brevity. Because 
the right side of Eq.\ (A8) is negative in $I$, these solutions are decreasing
functions in $I$. Also, if $F_1$ and $F_2$ are two solutions in $I$ with
$F_2(0) > F_1(0)$, then the difference $F_2-F_1$ is a finite, increasing
function in $I$ because it satisfies
$$(F_2 - F_1)'' =  \left({2\over 3}- 6u^2 \right)\left({F_2^2-F_1^2
\over F_1^2 F_2^2}\right) > 0.$$
Therefore solution curves do not cross in $I$, and the ``final value" $F(1/3)$
specifies a unique solution in $I$ (as does the initial value, $F(0)$). At
each $u\, \epsilon \, I$ the solutions in $I$ depend continuously (and 
monotonically) on initial and on final values. Now
consider the greatest lower bound (g.l.b.) of the final values. If this were 
positive it could be lowered, for example by integrating from a smaller
initial value, for the integration will run to $u = 1/3$ unless $F(u)$
approaches zero at some $u < 1/3$; but this cannot happen for finite final
values because $F(u)$ is a decreasing function. Hence the g.l.b. of final
values is zero. The corresponding limit of solutions $F(u)$ must therefore
be a solution that is positive in $0 \leq u < 1/3$ and approaches zero
at $u = 1/3$. This is the desired ``eigenfunction" that satisfies the
boundary condition (A9b).\footnote{The corresponding ``eigenvalue",
the g.l.b. of initial values $F(0) = \phi^2(0)$, can easily be found
numerically to be $0.3556...$.}

It remains to be verified that the solution satisfying the boundary conditions
in $u$ also satisfies the boundary condition in $x$. This follows from
Eq.\ (34b), $du/dx = -F(u)$. The asymptotic form of $F(u)$ near $u = \pm 1/3$, 
$$F(u) \rightarrow (12)^{1/3}\left({1\over 3} 
- |u|\right) \left[-\ln \left({1\over 3} - |u|\right)\right]^{1/3}$$
can be integrated to yield
$$ x \rightarrow \mp {3\over (96)^{1/3}} \left[-\ln \left({1\over 3} - |u|\right)
\right]^{2/3} + {\rm const}.$$
Thus $u = \mp 1/3$ does correspond to $x = \pm \infty$. (By deriving integral 
relationships from the differential equations the same conclusion, as well
as the fulfillment of the boundary conditions by all the unknown functions,
can be established without using asymptotic forms. We refrain from displaying
these somewhat involved relationships.)

\end{document}